# Symmetry energy and neutron-proton radii studies with a Wigner-Heisenberg monopole-monopole interaction

A.Z.Mekjian and L.Zamick

Department of Physics and Astronomy, Rutgers University, Piscataway, NJ 08854

Abstract

The symmetry energy in nuclei is studied using a monopole-monopole two boby interaction which has an isospin dependent term. A Hartree theory is developed for this interaction which has an oscillator shell model basis with corresponding shell structure. The role of shell structure on the symmetry energy is then studied. We also find that the strength of the Heisenberg interaction is very important for understanding the difference between proton and neutron radii and features associated with halo nuclei.

The symmetry energy appears in the Weizsacker mass formulae for the binding energy as a term that reflects the neutron excess. Specifically, the volume part has an $(N-Z)^2/A$ behavior with a coefficient $b_{sym} = 25 MeV$ when surface corrections to the symmetry energy are excluded. The mass number $A = N + Z$, with $N$ the number of neutrons and $Z$ the number of protons. The symmetry energy is important in understanding the valley of nuclear stability which involves the competition between the Coulomb energy which favors changing protons into neutrons and the symmetry energy which favors symmetric $N = Z$ nuclei. For fixed $A = N + Z$, the most stable isobar is determined by the minimum of $b_{sym}(A-2Z)^2/A + (3/5)Z^2 e^2/r_0 A^{1/3}$ or $Z/A = (1/2)/(1 + A^{2/3}(3/5)(e^2/r_0)/4b_{sym})$. Simple Fermi gas models [1] account for ½ the symmetry energy from kinetic energy considerations: $(b_{sym})_{kin} = (1/3)(e_f)_{N=Z} \approx 12 MeV$ where $e_f$ is the Fermi energy. The remaining ½ comes from interaction effects arising from the isopin dependence of the interaction. For example, in the independent particle approximation an isospin dependent one body Lane potential [2] of the particular form $V = V_0 + V_1 t_z (N-Z)/(2A)$, with $t_z = 1/2$ for neutrons and $t_z = -1/2$ for protons, can account for the interaction part of the symmetry energy. Summing this potential over protons and neutrons and including a double counting factor of ½ for two particle interactions gives $V_0 A/2 + V_1(1/8)(N-Z)^2/A$. A choice of $V_1 \approx 100 MeV$ accounts for the remaining part of the symmetry energy coefficient [1]. The $t_z (N-Z)/(2A)$ is the product of the third component of isospin $t_z$ of a single nucleon and the term $(N-Z)/(2A)$ is the third component of the isospin of the rest of the nucleons. A potential that is invariant under isospin space rotations involves the isospin vector product $\vec{t} \cdot \vec{T}$ resulting in $V = V_0 + (V_1/A)\vec{t} \cdot \vec{T}$. In this paper, we calculate the symmetry energy using a charge independent but isospin dependent two body interaction. In previous works[3,4] we did not have an isospin dependence in our interaction. Two reviews of properties of the symmetry potential are in ref. [5,6] and extended discussions appear in ref. [7,8]. The role of pairing correlations on the symmetry energy and incompressibility were recently presented in ref. [9]. The surface incompressibility was obtained by Sharma [10] in a relativistic model. Their has been a resurgence in this topic as in ref. [11,12]. In this work we stay strictly within the Hartree model [3] , but it should be noted that we have a separate paper which included correlations for symmetric nuclear matter [4]. Cluster correlations were considered in ref. [13].

Our two body interaction is

$$V(12) = \lambda r^2(1)r^2(2)[1 + b\vec{\tau}(1) \cdot \vec{\tau}(2)] \tag{1}$$

A monopole-monopole interaction $\lambda r^2(i)r^2(j)$ acting between nucleons leads to a one-body harmonic oscillator potential in a Hartree approximation. Particles can then be put into the lowest levels of this one-body potential to form a state given by a Slater determinate. In eq. (1), the $\vec{\tau} = \vec{t}$. The last factor $b\vec{\tau}(1) \cdot \vec{\tau}(2)$ has the value 1-3$b$ for two nucleons with total isospin $T = 0$ and the value 1+$b$ for two nucleons with total isospin $T = 1$. The isospin $T = 1$ are two protons or two neutrons or a neutron-proton pair couple to this value of isospin, while the $T = 0$ state is for a neutron-proton pair only. The np interaction is then ½ the sum or $1 - b$. Then $V(12, T = 0) = \lambda r^2(1)r^2(2)[1 - 3b]$ and $V(12, T = 1) = \lambda r^2(1)r^2(2)[1 + 3b]$. For positive $\lambda, b$ the $T = 1$ is more repulsive and the $T = 0$ more attractive. The number of $T = 0$ pairs is

$$N_{T=0} = \frac{A^2}{8} + \frac{A}{4} - \frac{T(T+1)}{2} \tag{2}$$

while the number of $T = 1$ pairs is

$$N_{T=1} = \frac{3A^2}{8} - \frac{3A}{4} + \frac{T(T+1)}{2} \tag{3}$$

The terms linear in mass number $A$ come from neutron and proton particles in the same space level each with the same spin $S_z$. These np pairs have to be coupled to isospin $T = 0$. The np pairs in the same space level with opposite spin $S_z$ are ½ $T = 0$ and ½ $T = 1$ states. The last remark also applies to all np pairs in different space states with opposite spin $S_z$ and also with the same or parallel spin $S_z$.

To determine a value for $b$ we consider $^{44}Ti$ and look at the energy difference between the $0^+, T = 0$ ground state and the $0^+, T = 2$ excited state at $9 MeV$. The $b\vec{\tau}(1) \cdot \vec{\tau}(2)$ interaction separates the two states by $12b$, pushing the $T = 2$ state up by $6b$ and pulling the $T = 0$ state down by $6b$. Therefore $b = 3/4$ since the separation is $9 MeV$.

We note that for a Heisenberg isospin exchange force $b = 1$ while a Wigner force $b = 0$ [14] in the expression $[1 + b\vec{\tau}(1) \cdot \vec{\tau}(2)]$. We allow $b$ to vary between 0 and 1. To proceed to evaluate the total energy, we first define $\Sigma_0$ as the sum of $2n + l + 3/2$ over all occupied proton states and neutron states for a nucleus of $A$ nucleons with equal numbers of protons and neutrons or $N_0 = Z_0 = (N+Z)/2$. The factor $2n + l + 3/2$ appears in single particle orbits of a harmonic oscillator potential with quantum numbers $n, l$ in polar coordinates which have energies given by $\varepsilon_{n,l} = (2n + l + 3/2)\hbar\omega$. In Cartesian coordinates this reads $\varepsilon_N = (N + 3/2)\hbar\omega$ with $N = n_x + n_y + n_z$. The degeneracy, including spin degeneracy $g_s = 2$, is $(N+1)(N+2)$. Then a sigma sum for a neutron or for a proton can be written as

$$\sum_{N=0}^{N_{max}}(N+\frac{3}{2})(N+1)(N+2) = \frac{1}{4}(N_{max}+2)^2(N_{max}+1)(N_{max}+3) \approx \frac{1}{4}(N_{max}+2)^4 \quad (4)$$

The $N_{max}$ is the $N$ of the last shell in the oscillator which is taken as filled to obtain this result. The sum of the degeneracy factor is

$$\sum_{N=0}^{N_{max}}(N+1)(N+2) = \frac{1}{3}(N_{max}+1)(N_{max}+2)(N_{max}+3) \approx \frac{1}{3}(N_{max}+2)^3 \quad (5)$$

Thus for filled major shells of a large number of protons $(N_{max}+2)^3/3 = Z$ and $(N_{max}+2)^4/4 = \Sigma_p$ giving $\Sigma_p = 3^{4/3}Z^{4/3}/4 \approx Z^{4/3}$. A similar result follows for a large number of neutrons. A more accurate result is $\Sigma_p = 3^{4/3}Z^{4/3}/4 + Z^{2/3}/4 \cdot 3^{1/3}$. Closed major shells in this simple oscillator model occur at $N_{max}$ equal to integer values starting at zero and the degeneracy of each level of one type of particle including spin is 2,6,12,20,30,42,56 for $N_{max}$ =0,1,2,3,4,5,6 respectively. The total number of particles that fill all the levels up to $N_{max}$ is the sum and is 2,8,20,40,70, 112,168 for the above cases. The corresponding value of the $\Sigma_p$ or $\Sigma_n$ values are 3,18,60, 150,315,588,1008. The approximation $\Sigma_p = 3^{4/3}Z^{4/3}/4$ is 312 for $Z = 70$ compared to 315 and $\Sigma_n \approx 1.0817N^{4/3}$ is 584 compared to 588 for $N = 112$. We will consider an $Z = 70$, $N = 112$, $A = 182$ nucleus below. We note that if we consider a symmetric nucleus $Z = N = 91$, $A = 182$ the value of $\Sigma_p + \Sigma_n$ =315+588=903 of the $Z = 70$ and $N = 112$ is the same as the $\Sigma_p + \Sigma_n$ system with $Z = N = A/2 = 91$. This is because we keep all particles in the same major shell. We will also consider a much larger hypothetical system with $Z = 140$ and $N = 190$ with that does not have this feature. Specically, $\Sigma_p + \Sigma_n$ =798+1205=2003, while the $Z = 165$ and $N = 165$ nucleus has $\Sigma_p + \Sigma_n \equiv \Sigma_0$ =985.5+985.5=1971. We note that away from closed major shells, the above result $\Sigma_p = 3^{4/3}Z^{4/3}/4 + Z^{2/3}/4 \cdot 3^{1/3}$ are good but in differences of large numbers involving $\Sigma's$ that will appear in results given below large errors will arise.

The scaling relation in the Fermi gas for the kinetic energy is $Z^{5/3}$ for protons and

$N^{5/3}$ for neutrons and involves a higher power of $Z$ and $N$. The lower power of $Z$ and $N$ for the oscillator is a result of the higher degeneracy in the oscillator. The $(N-Z)^2/A$ dependence of the kinetic energy for the Fermi gas arises from the expansion of $(A/2)^{5/3}((1-2T_z/A)^{5/3}+(1+2T_z/A)^{5/3}) \approx (A/2)^{5/3}2[1+(5/3)(2/3)(1/2)(2T_z/A)^2]$. By comparison $Z^{4/3}+N^{4/3} \approx [1+(4/3)(1/3)(1/2)(2T_z/A)^2]$ for the oscillator. Thus the numerical coefficient of $(2T_z/A)^2$ in each of the square brackets goes from 5/9 for the Fermi gas to 2/9 for the oscillator. From the remarks made at the end of the previous paragraph we note that the kinetic energy does not change when we change neutrons into protons in the same major shell. Thus shell corrections are large and can lead to much smaller kinetic energy departures from $Z \neq N$ systems to $Z = N$ systems with the same $A$.

We choose the interaction strength $\lambda$ as in a previous works [3,4] so that the Hartree energy is a minimum at oscillator energy $\hbar\omega_0$ for a nucleus with no neutron excess. This leads to the condition $\lambda b = (\hbar\omega_0)^2/(2\Sigma_0)$. We define $x_p = \omega_0/\omega_p$ and similarly $x_n = \omega_0/\omega_n$. The expression for the energy is then

$$\frac{E}{\hbar\omega_0} = \frac{1}{4\Sigma_0}\left(\Sigma_p^2 x_p^2 + \Sigma_n^2 x_n^2\right)(1+b) + \frac{1}{2\Sigma_0}\left(\Sigma_p \Sigma_n x_p x_n\right)(1-b) + \frac{\Sigma_p}{2x_p} + \frac{\Sigma_n}{2x_n} \qquad (6)$$

In eq. (6) the kinetic energy contributions to the energy $E/\hbar\omega_0$ are the last two terms. Let us first consider the case $x_p = x_n = x$ where the neutrons and protons are in the same well shape. Defining a neutron excess function as $\Delta = \Sigma_n - \Sigma_p$ and $\Sigma = \Sigma_n + \Sigma_p$

$$\frac{E}{\hbar\omega_0} = \frac{\Sigma^2}{\Sigma_0}\frac{x^2}{4} + \frac{\Sigma}{2x} + b\frac{\Delta^2}{\Sigma_0}\frac{x^2}{4} \qquad (7)$$

For $b = 0$, we get the previous results of the isospin independent interaction of Ref. [3,4]. For $Z = 70$ and $N = 112$, the $\Sigma = \Sigma_n + \Sigma_p = 903 = \Sigma_0$ and $\Delta = 588 - 315$. At $x = 1$ and $b = 0.1$ the $E/\hbar\omega_0 = 679.31$. This number will be compared to a calculation given below which minimizes the energy of Eq. 6. The symmetry energy is defined as the difference of this last result and the corresponding result for $Z = N = A/2$ where $\Sigma \to \Sigma_0, \Delta \to 0$.

$$\frac{E_{sym}}{\hbar\omega_0} = \frac{\Sigma^2 - \Sigma_0^2}{\Sigma_0}\frac{x^2}{4} + \frac{\Sigma - \Sigma_0}{2x} + b\frac{\Delta^2}{\Sigma_0}\frac{x^2}{4} \qquad (8)$$

For the case where the proton major shell is filled and the excess neutrons occupy a fraction or all the next major shell, then $\Sigma = \Sigma_0$ and the symmetry energy is

$$\frac{E_{sym}}{\hbar\omega_0} = b\frac{\Delta^2}{\Sigma_0}\frac{x^2}{4} \qquad (9)$$

The
$$\Delta^2 = (\Sigma_n - \Sigma_p)^2 \approx 1.0817^2(N^{4/3} - Z^{4/3})^2 = 1.0817^2((A/2+T_z)^{4/3} - (A/2+T_z)^{4/3})^2 =$$
$$1.0817^2(A/2)^{8/3}2^2(4/3)^2((N-Z)^2/A^2) \text{ and } \Sigma = \Sigma_0 = 2\Sigma_{p=n=A/2} \approx 2\cdot 1.0817(A/2)^{4/3}$$
giving

$$\frac{E_{sym}}{\hbar\omega_0} = b\frac{\Delta^2}{\Sigma_0}\frac{x^2}{4} \approx bx^2 1.0817\frac{8}{9}\frac{A^{1/3}}{2^{4/3}}\frac{(N-Z)^2}{A} \qquad (10)$$

Since $\hbar\omega_0 \approx 41 MeV / A^{1/3}$ the $E_{sym}$ has the well-known $(N-Z)^2/A$ behavior. We now consider the case where $x_p = \omega_0/\omega_p$ and $x_n = \omega_0/\omega_n$ are different. In the case the kinetic energy will contribute since the kinetic energy of a neutron and proton with the same quantum numbers is different. The energy of eq. (6) depends on these two variables and leads to two conditions by minimizing with respect to $x_p$ and $x_n$ which are

$$[\Sigma_p x_p(1+b) + \Sigma_n x_n(1-b)] - \Sigma_0/x_p^2 = 0 \qquad (11)$$

and

$$[\Sigma_n x_n(1+b) + \Sigma_p x_p(1-b)] - \Sigma_0/x_n^2 = 0 \qquad (12)$$

As a starting point, for the case considered above with $\Sigma_p = 315, \Sigma_n = 588, \Sigma_0 = 903$ and $b = 0.1$, the $x_n = 0.987187$ and the $x_p = 1.01615$. A three dimensional plot of the energy as a function of $x_p$ and $x_n$ is shown in Fig.1.

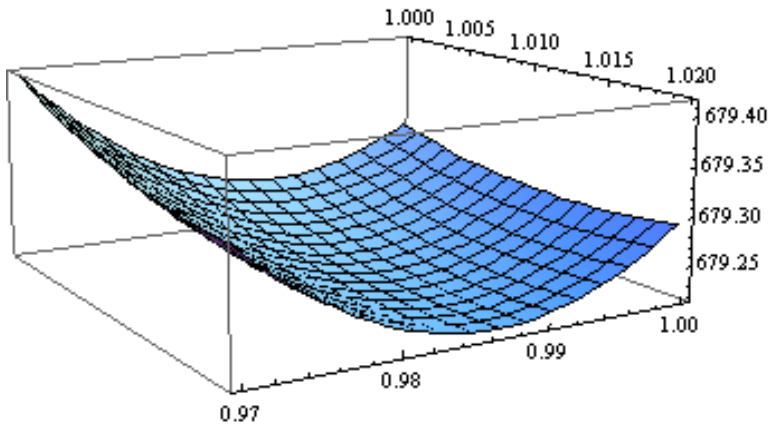

FIG.1. (Color online) The energy as a function of $x_p$ and $x_n$. The 3D plot is the energy of Eq. 6 for $\Sigma_p = 315, \Sigma_n = 588, \Sigma_0 = 903$. The $x_p$ minimum occurs at $x_p = 1.016$ and the $x_p$ axis runs from 1 to 1.2 while the $x_n$ minimum occurs at $x_n = .987$ and the $x_n$ axis runs from .97 to 1.0. The minimium energy is 679.22 which is very close to the approximate number of 679.31 from Eq. 7.

The behavior of $x_p, x_n$ with $b$ for $Z = 70, N = 112$ can be approximated as $x_p \approx 1 + .15b, x_n \approx 1 - .125b$ for small $b \leq 0.5$. The kinetic energy contribution to the symmetry energy arising from the difference in well shapes between protons and neutrons is then

$$\frac{E_{sym,kin}}{\hbar\omega_0} = \frac{\Sigma_p}{2x_p} + \frac{\Sigma_n}{2x_n} - \frac{\Sigma_0}{2} \approx 13b \qquad (13)$$

Let us consider next the behavior of the case were neutrons shell occupancy versus proton shell occupancy differ by two shells. We take $Z = 140, N = 190$ which has $\Sigma_p = 798, \Sigma_n = 1205, \Sigma_0 = 1971$. The exact solutions to the coupled equations for $x_p, x_n$ can be approximated by $x_p \approx 0.9946 + 0.09b$, $x_n \approx 0.9946 - 0.08b$ leading to

$$\frac{E_{sym,kin}}{\hbar\omega_0} = \frac{\Sigma_p}{2x_p} + \frac{\Sigma_n}{2x_n} - \frac{\Sigma_0}{2} \approx 16 + 5.43 + 12b \qquad (14)$$

The additional term of 16 comes from the $\Sigma_p + \Sigma_n - \Sigma_0 = 32$ and represents the contribution from the fact that there are two major shells in the difference of occupancy of protons from neutrons. Moreover, the solution at $b = 0$ is $x_p = x_n = .9946$, and no longer $x_p = x_n = 1$ which occurs when $\Sigma_p + \Sigma_n - \Sigma_0 = 0$. The factor 5.43 comes from this shift of $x_p = x_n = .9946$ away from 1.

The isospin dependent interaction of Eq. 1 moreover also influences the mean square radius of protons and neutrons are

$$\frac{<r_p^2>}{b_0^2} = x_p \frac{\Sigma_p}{Z}, \quad \frac{<r_n^2>}{b_0^2} = x_n \frac{\Sigma_n}{N} \qquad (15)$$

Table 1 gives the dependence of these mean square radii with interaction strength $b$. The results show that the initial difference between the neutron and proton mean square radii decreases with increasing strength $b$. Moreover at large values of the interaction strength the mean square radius of protons is larger than the mean square radius of neutrons. The lower half of the table includes a Coulomb energy.

Table 1. Behavior of the mean square radii with interaction strength $b$. The first case considered in the top part of the table is $Z = 140, \Sigma_p = 798, N = 190, \Sigma_n = 1205, \Sigma_0 = 1971$. The second case considered in the lower half of the table, which also includes a Coulomb term, is for $Z = 70, \Sigma_p = 315, N = 112, \Sigma_n = 588, \Sigma_0 = 903$.

| $b$ | $x_n$ | $x_p$ | $<r_n^2>/b_0^2$ | $<r_p^2>/b_0^2$ |
|---|---|---|---|---|
| Case 1. $Z = 140, \Sigma_p = 798, N = 190, \Sigma_n = 1205, \Sigma_0 = 1971$ and with no Coulomb | | | | |
| 0.1 | 0.9857 | 1.0050 | 6.252 | 5.729 |
| 0.2 | 0.9778 | 1.0147 | 6.201 | 5.784 |
| 0.5 | 0.9582 | 1.0400 | 6.077 | 5.928 |
| 0.6 | 0.9528 | 1.0474 | 6.043 | 5.970 |
| 0.75 | 0.9455 | 1.0577 | 5.996 | 6.029 |
| 1.00 | 0.9352 | 1.0728 | 5.931 | 6.115 |
| Case 2. $Z = 70, \Sigma_p = 315, N = 112, \Sigma_n = 588, \Sigma_0 = 903$ and with Coulomb | | | | |
| 0.0 | 0.9971 | 1.0221 | 5.23 | 4.60 |
| 0.1 | 0.9847 | 1.0385 | 5.17 | 4.67 |
| 0.2 | 0.9737 | 1.0532 | 5.11 | 4.74 |
| 0.3 | 0.9638 | 1.0675 | 5.06 | 4.80 |
| 0.4 | 0.9549 | 1.0806 | 5.01 | 4.86 |
| 0.5 | 0.9468 | 1.0937 | 4.97 | 4.92 |
| 0.6 | 0.9395 | 1.1057 | 4.93 | 4.98 |
| 0.7 | 0.9328 | 1.1167 | 4.90 | 5.03 |
| 0.8 | 0.9266 | 1.1272 | 4.86 | 5.07 |
| 0.9 | 0.9209 | 1.1372 | 4.83 | 5.12 |

The Coulomb energy can also be included approximately. We treat the one body potential as a uniform sphere and take the interior behavior of

$$V_C = \frac{3}{2}\frac{Ze^2}{R}(1 - \frac{1}{3}\frac{r^2}{R^2}) \tag{16}$$

for all $r$. The Coulomb energy in the $n, l$ level, which has $\varepsilon_{n,l}/2 = (2n + l + 3/2)\hbar\omega/2 = (1/2)m\omega^2 <r^2>_{n,l}$, is then

$$<V_C>_{nl} = \frac{3}{2}\frac{Ze^2}{R}(1 - \frac{1}{3}\frac{(\hbar/m\omega)(2n + l + 3/2)}{R^2}) \tag{17}$$

Taking ½ the sum over all occupied proton orbitals gives

$$E_C = \frac{3}{4}\frac{Z^2 e^2}{R} - \frac{1}{4}\frac{Ze^2}{R^3}\frac{\hbar}{m\omega}\Sigma_p \tag{18}$$

Since $R^2 = (5/3)(\hbar/m\omega)\Sigma_p / Z$, the resulting Coulomb energy is the well known result of a uniform density charged sphere which is $E_C = (3/5)Z^2 e^2 / R$. Using $x_p = \omega_0 / \omega_p$ and the above result for $R$, the energy of Eq. 6 becomes

$$\frac{E}{\hbar\omega_0} = \frac{1}{4\Sigma_0}\left(\Sigma_p^2 x_p^2 + \Sigma_n^2 x_n^2\right)(1 + b) + \frac{1}{2\Sigma_0}\left(\Sigma_p \Sigma_n x_p x_n\right)(1 - b) + \frac{\Sigma_p}{2x_p} + \frac{\Sigma_n}{2x_n} + \frac{K_C Z^{5/2}}{x_p^{1/2}\Sigma_p^{1/2}} \tag{19}$$

The Coulomb constant $K_C = (3/5)^{3/2}(e^2/\hbar c)/(\hbar\omega_0/mc^2)^{1/2} \approx 0.04$ at $A = 182$. The minimization condition of Eq. 12 is unchanged, while that of Eq. 11 is changed to

$$[\Sigma_p x_p(1 + b) + \Sigma_n x_n(1 - b)] - \Sigma_0 / x_p^2 - K_C Z^{5/2} /(2\Sigma_p^{1/2} x_p^{3/2}) = 0 \tag{20}$$

The last or Coulomb term at $Z = 70, A = 182, x_p = 1$ is ~90 which can be compared to kinetic energy term of $\Sigma_0 / x_p^2$ ~900 at $x_p = 1$. The $x_p, x_n$ minima occur at various values determined by the strength of the $\vec{\tau}(1) \cdot \vec{\tau}(2)$ interaction as given in the above table.

    In this paper we studied properties of the symmetry energy in a Hartree model. A monopole-monopole interaction which includes an isospin-dependent but charge-independent interaction is used. Specifically, we included a term $[1 + b\vec{\tau}(1) \cdot \vec{\tau}(2)]$ where $b = 1$ for a Heisenberg isospin exchange force while $b = 0$ for a Wigner force. We allowed $b$ to vary between 0 and 1. The monopole-monopole interaction leads to an oscillator shell model basis with corresponding shell structure. The results are compared to a Fermi gas model to study the role of shell structure on the symmetry energy. We found that shell structure can play an important role in the magnitude of the symmetry energy. This is easily seen in the kinetic energy contribution since changing a neutron into a proton in the same shell has very little effect on the kinetic energy contribution if the oscillator frequency of protons and neutrons are the same. The kinetic energy contribution is about ½ the total symmetry energy in a Fermi gas model [1]. The oscillator well parameters, defined as $x_p, x_n$, change in our Hartree model as a function of the strength of the isospin dependent part of the interaction. However, the change in kinetic energy due to the change in the oscillator parameters is found to be small when changing neutrons into protons in the same shell. Correspondingly, the mean square radius of protons and neutrons change with changes in $x_p, x_n$. The results of table 1 show that the initial difference between the neutron and proton mean square radii decreases

with increasing strength $b$. Moreover at large values of the interaction strength the mean square radius of protons is larger than the mean square radius of neutrons. Thus we find that correlations are very important for understanding the difference between proton and neutron radii which, in turn, has consequences for halo nuclei.

Acknowledgements. Supported in part by a DOE grant DE-FG02-96ER-4097.

References
1. A A.Bohr and B.R.Mottelson, Nuclear Structure, Vol.I, (W.A.Benjamin, NY 1969)
2. A.M.Lane, Nucl. Phys. 35, 676 (1962)
3. L.Zamick, Nucl. Phys. A232, 13 (1974)
4. A.Z.Mekjian and L.Zamick, arXiv:1112.0457 nucl-th
5. A.W.Steiner, M.Prakash, J.M.Lattimer, P.J.Ellis, Phys. Rep. 411, 325 (2005)
6. B.A.Li, Lie-Wen Chen, Che Ming Ko, Phys. Rep. 464, 113 (2008)
7. P.Danielewicz and J.Lee, Int. J. Mod. Phys. E18, 829 (2009)
8. P.Danielewicz and J.Lee, Nucl. Phys. A818, 36 (2009)
9. E. Khan, J. Margueron, G. Colo, K. Hagino, H. Sagawa, arXiv:1005.1741
10. M.M.Sharma, Nucl. Phys. A816, 65 (2009)
11. H. Jlang, G.J.Fu, M.Zhao and A.Arima, Phys. Rev. C85, 024301 (2012)
12. S.Shlomo, Journ of Phys., Conf Series 337, 012014 (2012)
13. A.Z.Mekjian, S.J.Lee, L.Zamick, Phys. Rev. C72, 044305 (2005)
14. J.M.Blatt and V.Weisskopf, Theoretical Nuclear Physics, (John Wiley & Sons, NY 1952)